\documentclass[prd,aps,showpacs,nofootinbib,preprint,eqsecnum]{revtex4}

\usepackage{amsmath}
\usepackage{amssymb}
\usepackage{amsfonts}
\usepackage{graphicx,bm}
\usepackage{dcolumn}

\usepackage{color,amsxtra}
\usepackage{epsf}

\usepackage{enumerate}
\usepackage{hhline}
\usepackage{array}
\usepackage{tabularx}

\usepackage{subfigure}

\usepackage{fancyhdr}
\usepackage{mathrsfs}

\usepackage{amsmath}
\usepackage{amsthm}
\usepackage{amssymb}
\usepackage{amscd}
\usepackage[british]{babel}
\usepackage{babel}
\usepackage{graphicx}
\usepackage{psfrag}
\usepackage{epsfig}
\usepackage{rotating}
\usepackage{times}

\theoremstyle{plain}

\newtheorem{remark}{Remark}[section]

\newcommand{\boxend}{\flushright{$\Box$}}

\renewcommand{\tilde}{\widetilde}

\begin{document}


\title{On $R+\alpha R^2$ Loop Quantum Cosmology}
\author{
Jaume Amor\'os$^{1, }$\footnote{E-mail address: jaume.amoros@upc.edu},
Jaume de Haro$^{1, }$\footnote{E-mail address: jaime.haro@upc.edu},
and
Sergei D. Odintsov$^{2, 3, 4,  }$\footnote{
E-mail address: odintsov@ieec.uab.es}
}


\affiliation{
$^1$Departament de Matem\`atica Aplicada I, Universitat
Polit\`ecnica de Catalunya, Diagonal 647, 08028 Barcelona, Spain \\
$^2$Instituci\`{o} Catalana de Recerca i Estudis Avan\c{c}ats (ICREA),
Barcelona, Spain\\
$^3$Institut de Ciencies de l'Espai (CSIC-IEEC),
Campus UAB, Facultat de Ciencies, Torre C5-Par-2a pl, E-08193 Bellaterra
(Barcelona), Spain\\
$^4$Tomsk State Pedagogical University, 634061 Tomsk and National Research Tomsk State University, 634050 Tomsk, Russia
}



\begin{abstract}
Working in Einstein frame
we introduce, in order to avoid singularities, holonomy corrections to the  $f(R)=R+\alpha R^2$ model.  We perform a detailed analytical and numerical study  when
holonomy corrections are taken into account in both Jordan and Einstein frames obtaining, in Jordan frame, a  dynamics which differs qualitatively, at early  times, from the one
of the original model. More precisely, when holonomy corrections are taken into account the universe is not singular,
starting at early times in the contracting phase and  bouncing to enter in the expanding one
where, as in the original model, it inflates. This dynamics is  completely different  from the one obtained in the original $R+\alpha R^2$ model, where the universe is singular at
early times and never
bounces. Moreover, we show that these holonomy corrections may lead to better predictions for the inflationary phase as compared with current observations.
\end{abstract}

\pacs{
04.60.Pp, 98.80.Jk, 04.50.Kd
\\
{\footnotesize Keywords:
Loop quantum cosmology;\
Dynamical systems;\ Modified gravity.}}

\maketitle



\section{Introduction}

Two kind of
quantum geometric corrections come  from the discrete nature of space-time assumed in Loop Quantum Cosmology (LQC): inverse volume corrections \cite{bojowald}
and holonomy corrections (see for instance \cite{ashtekar}). Dealing with the flat Friedmann-Lem\^itre-Robertson-Walker (FLRW) geometry, which is the case of our paper,
inverse volume corrections have 
problems because of arbitrary re-scalings \cite{as11}, more precisely,  
since the scale factor can be arbitrarily re-scaled in a flat metric, these
inverse volume correction could appear an any arbitrary scale loosing its physical meaning. Only in a closed universe they have sense,  leading to a bounce that avoids
the big bang and big crunch singularity \cite{toporensky}.
 On the other hand, holonomy corrections, which are well introduced for compact and no-compact geometries,  provide a big bounce that avoids singularities like the Big Bang and Big Rip
 (see for example \cite{singh}).

On the other side, it is well-known that, in general, f(R) gravity does not avoid singularities, except of particular non-singular cases where $R^2$ term plays an important role as it was demonstrated in
 \cite{bno08}. In order to avoid them, one could
introduce
holonomy corrections in $f(R)$ gravity. The extension of Loop Quantum Gravity (LQG) to $f(R)$ gravity   has  been recently developed in \cite{zm11,zm11a}, where
holonomy corrections  are introduced in Einstein frame (EF), because in that frame
 the gravitational
part of the Hamiltonian is linear in the scalar curvature and the matter part is given by a scalar field. It is important to recall that, the idea to introduce holonomy corrections
via the EF was performed in \cite{Risi} studying the gracefull exit problem in the pre-big bang scenario, i.e., studying the regularization of the singularity that divides the pre
and post big bang branches in pre-big bang models.

This extension simplifies very much
when one consider the flat FLRW geometry. In that case,
in order
to take into account geometric effects, one has to replace the Ashtekar connection by a suitable sinus function (see for instance
\cite{h13a}) obtaining the holonomy corrected Friedmann equation in EF.
Finally,
from the holonomy corrected Friedmann equation in EF and through the relation between the corresponding variables in both frames, one obtains the holonomy corrected
$f(R)$ theory in the Jordan frame (JF).

Our main objective is to apply, for the flat FLRW geometry, holonomy correction to the  $f(R)=R+\alpha R^2$ model (also called $R^2$ gravity)
and study its dynamics. To do this, first of all we perform
 a detailed analysis of $R^2$ gravity without corrections. When
holonomy corrections in the model are taken into account one  obtains a very  complicated dynamical equation in the JF. Fortunately, dynamical equations simplify very much in
EF, (in fact the dynamics is given by  the well-known holonomy corrected Friedmann equation in LQC
plus the Klein-Gordon equation in flat FLRW geometry)
which allows us to perform a very deep analytical an numerical analysis, whose results can be translated to the JF.
Our conclusion is that when holonomy corrections are taken into account the universe starts at the critical point $(H=0,\dot{H}=0)$ (the Hubble parameter and its derivative  vanish)
and makes small oscillations around the critical point before entering the contracting phase, which it leaves bouncing
(see  \cite{brandenberger} for a review of bounce cosmology), and enters the expanding phase where, as in the
classical model, it
reaches an inflationary stage which it leaves
at late times and comes back to  the critical point, once again,  in an oscillating way.

The paper is organized as follows:

In Section II, we review $f(R)$ gravity in Jordan and Einstein frames.
In Section III, we introduce holonomy corrections to $f(R)$ gravity. The idea is very simple: working in EF,  $f(R)$ gravity is formulated as Einstein gravity plus and scalar field.
Then, the idea, as in standard LQC for the flat FLRW geometry, is to replace the Ashtekar connection by a suitable sinus function.
Section IV is devoted to the study of $R^2$ gravity without holonomy corrections. After performing the change of variable $p^2=H$ where $H$ is
the Hubble parameter,
the obtained dynamical equation  can be understood as the dynamics of a particle under the action of a quadratic potential with dissipation. This system is very simple and the phase portrait
can be drawn with all the details.
In Section V, we analyze the model with holonomy corrections. We start working in EF due to the simplicity of equations and, once we have studied the dynamics in EF,
we  obtain the dynamics in JF
form the formulae that relate both frames.
Moreover, we obtain in EF the corrected expressions of the slow-roll parameters and the values of the spectral index for scalar perturbations and the ratio of tensor to scalar
perturbations, showing that holonomy corrections help to match correctly the theoretical results obtained from $R^2$ gravity with current observations.
Section VI is devoted to discuss a possible unification of inflation and current cosmic acceleration in  the framework of Loop Quantum $f(R)$ theories. We will show that
 when one consider the current suggested models for such unification
this extension and/or its analytical study is, in general, unworkable. The only model we have been able to deal with is $R^2$ plus an small cosmological constant. For such a model, we have
performed a detailed analytical
study and the results are shown at the end of the work.
\section{Classical dynamical equations in different frames}

In this Section we review the relations between Jordan and Einstein frames in $f(R)$ gravity for the flat FLRW geometry.

 The  Lagrangian in JF for the flat FLRW geometry is given by ${\mathcal L}_{JF}=\frac{a^3}{2}f(R)$, where the scalar curvature is
 $R=6\dot{H}+12H^2$ being $H=\frac{\dot{a}}{a}$  the Hubble parameter, and
the corresponding modified Friedmann equation in $f(R)$ gravity can be obtained form Ostrogradskii’s construction  \cite{h13a}  giving as a result
\begin{eqnarray}\label{1}
 6f_{RR}(R)\dot{R}H+(6H^2-R)f_R(R)+f(R)=0,
\end{eqnarray}
 where $f_R(R)\equiv \frac{\partial f(R)}{\partial R}$.
Taking the derivative of equation (\ref{1}) with respect to time and using the relation $R=6(\dot{H}+2H^2)$ one obtains the equivalent equation
 \begin{eqnarray}\label{1a}
 f_{RR}(R)(\ddot{R}-\dot{R}H)+f_{RRR}(R)\dot{R}^2+2f_R(R)\left(\frac{R}{2}-2H^2\right)=0.
\end{eqnarray}

To work in the Einstein frame (EF), one has to perform the change of variables \cite{dsy}
\begin{eqnarray}\label{2}
\tilde{a}=\sqrt{f_R(R)}a; \quad d\tilde{t}=\sqrt{f_R(R)}dt.
\end{eqnarray}

Then, in that frame the Lagrangian density, for flat FLRW geometries, is
\begin{eqnarray}\label{3}
{\mathcal L}_{EF}=\tilde{a}^3\left(\frac{1}{2}\tilde{R}+\frac{1}{2}(\tilde{\phi}')^2-V(\tilde{\phi})\right)
\Longleftrightarrow {\mathcal L}_{EF}=-3(\tilde{a}')^2\tilde{a}+\tilde{a}^3\left(\frac{1}{2}(\tilde{\phi}')^2-V(\tilde{\phi})\right),
\end{eqnarray}
where $'$ means the derivative with respect the time $\tilde{t}$. Here,
$\tilde{a}$ and $\tilde{\phi}$ have to be considered as independent variables, and of course, $\tilde{R}=6{\tilde{H}}'+12\tilde{H}^2$.

The relation between both frames is given through the relations
\begin{eqnarray}\label{4}
 \tilde{\phi}=\sqrt{\frac{3}{2}}\ln(f_R(R));\quad V(\tilde{\phi})=\frac{Rf_R(R)-f(R)}{2f_R^2(R)},
\end{eqnarray}
and
a simple calculation shows that the Friedmann equation in the EF, i.e. $\tilde{H}^2=\frac{1}{3}\tilde{\rho}$,
  obtained from the Hamiltonian constrain
\begin{eqnarray}\label{ham}{\mathcal H}_{EF}\equiv\tilde{a}'\frac{\partial {\mathcal L}_{EF}}{\partial\tilde{a}'}+
\tilde{\phi}'\frac{\partial {\mathcal L}_{EF}}{\partial\tilde{\phi}'}-{\mathcal L}_{EF}=
-3(\tilde{a}')^2\tilde{a}+\tilde{a}^3\left(\frac{1}{2}(\tilde{\phi}')^2+V(\tilde{\phi})\right)=0,
\end{eqnarray} where
$\tilde{\rho}\equiv
\frac{1}{2}(\tilde{\phi}')^2+V(\tilde{\phi})$,
is equivalent to equation (\ref{1}).
However, the Friedmann equation in EF, $\tilde{H}^2=\frac{1}{3}\tilde{\rho}$, is a constrain instead of a dynamical equation. The dynamics  is given by  the conservation
equation $\tilde{\rho}'=-3\tilde{H}(\tilde{\phi}')^2$ or the Raychauduri one $\tilde{H}'=-\frac{1}{2}(\tilde{\phi}')^2$ which are equivalent  to equation (\ref{1a}).

Note that combining, in EF, the conservation and Friedmann equation one obtains
\begin{eqnarray}
 (\tilde{\rho}')^2=3\tilde{\rho}(\tilde{\phi}')^2,
\end{eqnarray}
and coming back to the JF this equation is a second order differential equation in $R$ (it only contains $R$, $\dot{R}$ and $\ddot{R}$) which is equivalent to equations
(\ref{1}) and (\ref{1a}).

Finally, we show
the following relations between both frames, which will be important when we extend LQC to $R^2$ gravity:
\begin{eqnarray}\label{rel}
 {H}=\sqrt{f_R(R)}\left(\tilde{H}-\frac{1}{\sqrt{6}}\tilde{\phi}' \right);\quad R=f_R(R)\left(\tilde{R}+(\tilde{\phi}')^2+
\sqrt{6}\frac{\partial V(\tilde{\phi})}{\partial\tilde{\phi}} \right).
\end{eqnarray}

\section{$f(R)$ Loop Quantum Cosmology}
The idea to extend Loop Quantum Cosmology to $f(R)$ theories  ($f(R)$ LQC) has been recently developed in \cite{zm11,zm11a}.  For a flat FLRW geometry the idea is very simple and
 goes as follows: Working in EF, in the same way as in standard LQC,
we can see that the classical variable $\tilde{\beta}\equiv\gamma\tilde{H}$, where $\gamma$ is the Barbero-Immirzi parameter,  and the
volume $\tilde{V}\equiv\tilde{a}^3$ are canonically conjugated variables with Poisson bracket $\{\tilde{\beta},\tilde{V}\}=\frac{\gamma}{2}$ \cite{singh09}.
Then, in order to take into account the discrete nature of the space, one has to choose  a Hilbert space
where quantum states was represented by  almost
periodic functions. However, in this space the variable $\tilde\beta$ does not correspond to a well-defined quantum operator in this space, and since it
appears in the Hamiltonian (\ref{ham}), because it could be written as
\begin{eqnarray}\label{ham1}{\mathcal H}_{EF}=
-3\frac{\tilde{\beta}^2}{\gamma^2}\tilde{V}+\tilde{V}\left(\frac{1}{2}(\tilde{\phi}')^2+V(\tilde{\phi})\right)=0,
\end{eqnarray}
in order to have a well-defined quantum theory one needs to use the general holonomy corrected Hamiltonian in Loop Quantum Gravity
(see for instance \cite{aps06, abl03}).

At effective level, this is equivalent
 to introduce the square root of the minimum eigenvalue of the area operator in LQG, namely
 $\lambda=\sqrt{\frac{\sqrt{3}}{2}\gamma}$,  and
 make the replacement (see \cite{he10,dmw09,bo08} for a detailed discussion about the justification of this replacement)
\begin{eqnarray}\label{rep}
 \tilde{\beta}\rightarrow \frac{\sin(\lambda\tilde{\beta})}{\lambda},
\end{eqnarray}
in the Hamiltonian (\ref{ham1}), while keeping on the Poisson bracket $\{\tilde{\beta},\tilde{V}\}=\frac{\gamma}{2}$.
Here is important to notice that, after the introduction of holonomy corrections, $\tilde{\beta}$ stops to be equal to $\gamma \tilde{H}$. This can be showed from
the Hamilton equation $\tilde{V}'=\{\tilde{V},{\mathcal H}_{EF,LQC}\}$
 (being ${\mathcal H}_{EF,LQC}$ the new Hamiltonian obtained from (\ref{ham1}) after the replacement (\ref{rep})), because this equation can be written as
 $\tilde{V}'=-\frac{\gamma}{2}\frac{\partial{\mathcal H}_{EF,LQC} }{\partial \tilde{\beta}}=3\frac{\sin \lambda\tilde{\beta}\cos \lambda\tilde{\beta}}{\lambda\gamma}$
which is equivalent to
\begin{eqnarray}
 \tilde{H}=\frac{\sin 2\lambda\tilde{\beta}}{2\lambda\gamma}\Longleftrightarrow \tilde{\beta}=\frac{1}{2\lambda}\arcsin 2\lambda\gamma \tilde{H}.
\end{eqnarray}

\begin{remark}
It is stated in \cite{corichi} that there are many different inequivalent loop quantization, i.e., different pairs of canonically
conjugated variables leading to inequivalent quantum realizations. Here we have used the so-called
new quantization of LQC (also known $\bar{\mu}$ quantization) \cite{aps06} based in the use of variables $(\tilde{\beta},\tilde{V})$,
which is the unique choice consistent with the physical requirements proposed in \cite{corichi}.
\end{remark}

\begin{remark}
 It is important to stress that the replacement (\ref{rep}) is only valid for spatially flat models which is our case. When the spatial curvature does not vanish holonomy corrections
 has to be introduced in a non-trivial way (see for instance \cite{Gupt})
\end{remark}

Finally, from the Hamilton equation $\tilde{V}'=\{\tilde{V},{\mathcal H}_{EF,LQC}\}$ and the Hamiltonian constrain
${\mathcal H}_{EF,LQC}=0$, one obtains the
corresponding holonomy corrected version of the classical Friedmann equation \cite{singh09}, that is,
\begin{eqnarray}\label{5}
\tilde{H}^2=\frac{1}{3}\tilde{\rho}\left(1-\frac{\tilde{\rho}}{\tilde{\rho}_c}\right),
\end{eqnarray}
where $\tilde{\rho}_c\equiv \frac{3}{\lambda^2\gamma^2}$ is the so-called {\it critical density} in the EF.

As has been discussed in detail in \cite{aho} this equation depicts an ellipse in the plane $(\tilde{H},\tilde{\rho})$, and the
dynamics along this curve is very simple: For a non-phantom field the universe moves  clockwise from the contracting to the expanding phase
starting and ending at the critical point  $(0,0)$ and
bouncing
only once at $(0,\tilde{\rho}_c)$.

Finally,
note that
in the JF, the holonomy corrected Friedmann equation acquires the complicated form
\begin{eqnarray}\label{fri}
  6f_{RR}(R)\dot{R}H+(6H^2-R)f_R(R)+f(R)=
-\frac{\left(\frac{3}{2}f_{RR}^2(R)\dot{R}^2+(Rf_R(R)-f(R))f_R(R)\right)^2}{2f_R^4(R)\tilde{\rho}_c}.
\end{eqnarray}

\section{$R^2$ gravity}
In this Section we study with all the details the classical  model $f(R)=R+\alpha R^2$, with $\alpha>0$.
This model contains a quadratic correction to the scalar curvature and is a modified version of the  Starobinsky model \cite{s80}, where the author considered quantum vacuum
effects due to  massless fields conformally coupled with gravity. Note that such (eternal) trace-anomaly driven inflation was proposed earlier in ref \cite{dc}.

For this model, the classical equation (\ref{1}) becomes
\begin{eqnarray}\label{sta}
 12\alpha H\dot{R}+6H^2+12\alpha RH^2-\alpha R^2=0\Longleftrightarrow H^2=-12\alpha\left(3\dot{H}H^2+H\ddot{H}-\frac{1}{2}\dot{H}^2\right),
\end{eqnarray}
which coincides, when the parameter $\beta$ vanishes, with the dynamical  equation studied in \cite{s80}
\begin{eqnarray}\label{sta1}
 H^2=-12\alpha\left(3\dot{H}H^2+H\ddot{H}-\frac{1}{2}\dot{H}^2\right)+\beta H^4, \quad \mbox{where}\quad \beta>0.
\end{eqnarray}


It is very simple to show that equation (\ref{sta}) leads to an inflationary epoch \cite{mms, no11}. Effectively, when the slow-roll initial condition
$|\dot{H}|\ll H^2$ is fulfilled, equation (\ref{sta}) becomes $\ddot{H}=-3H\dot{H}-\frac{H}{12\alpha}$, which has  the following particular solution
in the expanding phase ($H>0$)
\begin{eqnarray}
\dot{H}(t)=-\frac{1}{36\alpha}\Longrightarrow
 H(t)=\frac{t_1-t}{36\alpha}\Longrightarrow a(t)=a(t_1)e^{-18\alpha H^2(t)}\quad \mbox{for}\quad t<t_1.
\end{eqnarray}

If  $t_i$ and  $t_f$ are the beginning and the end of inflation ($t_i<t_f<t_1$), then one will have
\begin{eqnarray}
 a(t_f)=a(t_i)e^{18\alpha (H^2(t_i)-H^2(t_f))}\cong a(t_i)e^{18\alpha H^2(t_i)},
\end{eqnarray}
and  the $60$ e-folds needed to solve the flatness and horizon problems  will be obtained when $\alpha H^2(t_i)$ is approximately $3.3$.


Unfortunately, $R^2$  gravity contains singularities at early times, that is, all solutions have divergent scalar curvature at early times. To show that, one has to perform the change
of variables $p^2(t)=H(t)>0$ \cite{aw} (in this model the universe doesn't bounce), then equation (\ref{sta}), which is not well-defined at singular value $H=0$, becomes
the following well-defined equation
\begin{eqnarray}\label{sis}
 \frac{d}{dt}\left(\frac{\dot{p}^2}{2}+W(p) \right)=-3p^2\dot{p}^2,
\end{eqnarray}
where $W(p)=\frac{p^2}{48\alpha}$.

We can see that the system (\ref{sis}) is dissipative. To understand its dynamics, we can imagine a "particle" rolling down along the parabola $W(p)$ losing energy
and oscillating, at late times, around $p=0$. As a
consequence, when time goes back the "particle" gains energy and finally $|p|\rightarrow \infty$ ($H\rightarrow \infty$), i.e., all the solutions
are singular at early times.
One also can check this fact as follows: We write equation (\ref{sis}) as
\begin{eqnarray}\label{sis1}
 \ddot{p}+\frac{p}{24\alpha}=-3p^2\dot{p},
\end{eqnarray}
and look for, at early times, solutions of the form $p(t)=\frac{C}{(t-\bar{t})^r}$, where $C$ and $r$ are parameters. Inserting this expression in
(\ref{sis1}) and retaining the leading terms when $t\gtrsim \bar{t}$, one obtains the equation:
\begin{eqnarray}
 \frac{r(r+1)C}{(t-\bar{t})^{r+2}}= \frac{3rC^3}{(t-\bar{t})^{3r+1}},
\end{eqnarray}
which has singular solutions at $t=\bar{t}$ of the form $p(t)=\sqrt{\frac{1}{2(t-\bar{t})}}$.

\begin{remark}
 In the contracting phase we can perform the change of variable $p^2(t)=-H(t)>0$, obtaining the system
 \begin{eqnarray}\label{sis0}
 \frac{d}{dt}\left(\frac{\dot{p}^2}{2}+W(p) \right)=3p^2\dot{p}^2,
\end{eqnarray}
where $W(p)=\frac{p^2}{48\alpha}$. We can see that the in the contracting phase the system is anti-dissipative (the universe gains energy), in this case the universe starts oscillating
around the bottom of the potential leaving it gradually and becomes singular at late times.
\end{remark}

Equation (\ref{sis1}) is also useful to obtain the inflationary period and the dynamics at late times. Effectively,
 when initially one has $\ddot{p}\cong 0$ equation (\ref{sis1}) becomes $p\dot{p}=-\frac{1}{72\alpha}$, whose inflationary solution is once again
\begin{eqnarray}\label{infla}
 H(t)=p^2(t)=\frac{t_1-t}{36\alpha}.
\end{eqnarray}

On the other hand, to obtain the dynamics at late times we follow the same method used in chaotic inflation for a quadratic potential (see page $240$ of \cite{mukh}).  Performing the
change of variable
\begin{eqnarray}\label{newvar}
\dot{p}(t)=\sqrt{2}f(t)\cos(\theta(t)),\quad {p}(t)=\sqrt{48\alpha}f(t)\sin(\theta(t)),
\end{eqnarray}
and inserting these expressions in equations (\ref{sis}) and (\ref{sis1}) one gets the system
\begin{eqnarray}\label{sis2}\left\{\begin{array}{ccc}
 \dot{f}&=&-18\alpha f^3(1-\cos(4\theta))\\
  \dot{\theta}&=& \frac{1}{\sqrt{24\alpha}}+144\alpha f^2\sin^3(\theta)\cos(\theta).              \end{array}\right.
\end{eqnarray}

Since $p$ goes to zero at late times, we can disregard the second term in the right hand side in the second equation of (\ref{sis2}), obtaining
$\dot{\theta}=\frac{1}{\sqrt{24\alpha}}$, whose solution is $\theta(t)=\frac{t}{\sqrt{24\alpha}}+\omega$, $\omega$  being a constant of integration. Inserting this approximate solution in the
first equation of (\ref{sis2}), we obtain a solvable equation whose solution is given by
\begin{eqnarray}
 f(t)=\sqrt{\frac{1}{36\alpha t\left(1-\frac{\sin\left(\frac{2t}{\sqrt{6\alpha}}+4\omega\right)}{\frac{2t}{\sqrt{6\alpha}}}\right)}}
 \cong
 \sqrt{\frac{1}{36\alpha t}\left(1+\frac{\sin\left(\frac{2t}{\sqrt{6\alpha}}+4\omega\right)}{\frac{2t}{\sqrt{6\alpha}}}\right)},
\end{eqnarray}
and thus the Hubble parameter reads
\begin{eqnarray}
 H(t)\cong\frac{4}{3t}
 \left(1+\frac{\sin\left(\frac{2t}{\sqrt{6\alpha}}+4\omega\right)}{\frac{2t}{\sqrt{6\alpha}}
 }\right)\sin^2\left(\frac{t}{\sqrt{24\alpha}}+\omega\right).
\end{eqnarray}

Now,
choosing $\omega=\pi/2$ one obtains the well-known result \cite{mms,w,wa}
\begin{eqnarray}
 H(t)\cong\frac{4}{3t}
 \left(1+\frac{\sin\left(\frac{2t}{\sqrt{6\alpha}}\right)}{\frac{2t}{\sqrt{6\alpha}}
 }\right)\cos^2\left(\frac{t}{\sqrt{24\alpha}}\right),
\end{eqnarray}
and after integrating by parts  one gets as Starobinsky in \cite{s80}
\begin{eqnarray}\label{H3}
 a(t)\cong t^{2/3}\left(1+\frac{2}{3}\frac{\sin\left(\frac{t}{\sqrt{6\alpha}}\right)}{\frac{t}{\sqrt{6\alpha}}}\right)\cong t^{2/3}.
\end{eqnarray}

These analytic results are supported numerically in figure $1$:

\vspace{2cm}

\begin{figure}[h]
\begin{center}
\includegraphics[scale=0.4]{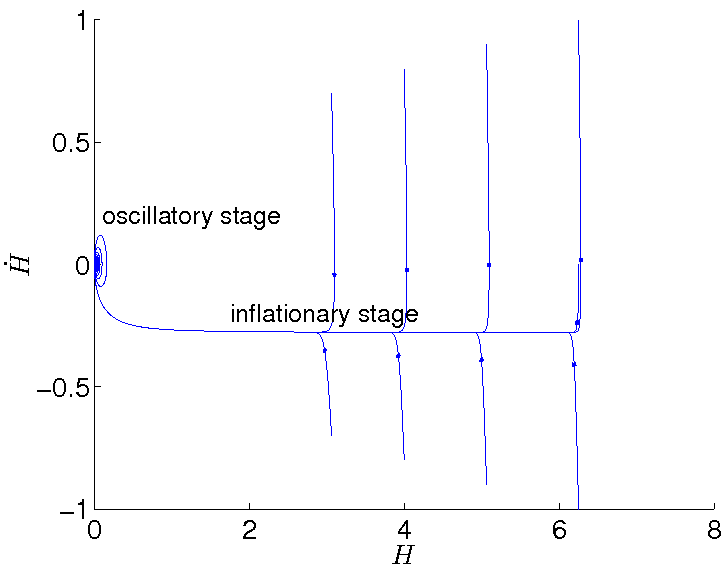}
\includegraphics[scale=0.4]{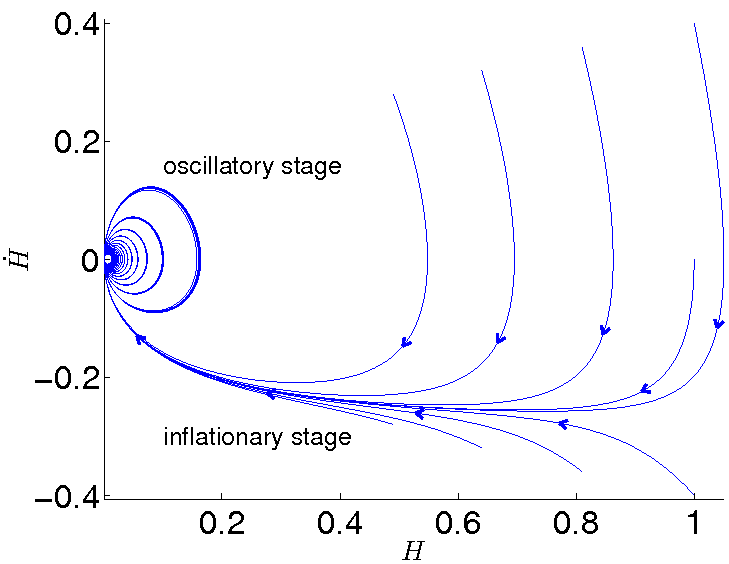}
\end{center}
\caption{{\protect\small Phase portrait for $\alpha=0.1$. The universe comes from a  singularity at early times, when time goes forward it enters in the attractor
inflationary phase, leaving it at early times
when the universe
starts to oscillate around $(0,0)$ without bouncing. In the first figure we have taken values of $H$ up to $6$ to show clearly the inflationary stage, and the second one
the values of $H$ are up to $1$ to show, in more detail, the oscillatory phase. It's clear from the pictures that orbits are unbounded, coming form $\infty$ at early times.}}
\end{figure}

An important remark is in order:
Note that in the  model of \cite{s80} one obtains the same equation (\ref{sis}) but with the potential  $W(p)=\frac{p^2}{48\alpha}-\frac{\beta p^6}{144\alpha}$.
In this case the potential has an stable minimum at $p=0$ and two unstable  maximums at $p=\pm\beta^{-1/4}$, which corresponds to the unstable de Sitter solution
$H=\beta^{-1/2}$. From the shape of this potential one deduces that
there are only two unstable non-singular solutions (the ones that start at the de Sitter points and end at the bottom of the potential), and two that only are
singular at late time (the ones that start at the de Sitter points and ends at $|p|=\infty$),
all the other solutions are singular at early times.  At late times, there are two kinds of solutions: the ones that
have enough energy to overpass  the wedge of the potential and become singular at late times, and others with less energy that fall down against the wedge of the potential without clearing
it  due to the dissipation and, approaching to $p=0$ with the same oscillatory behavior as in the $R+\alpha R^2$  model (see figure 2 for the shape of potentials, and figure 3
for the phase portrait of the Starobinsky model).

\begin{figure}[h]
\begin{center}
\includegraphics[scale=0.3]{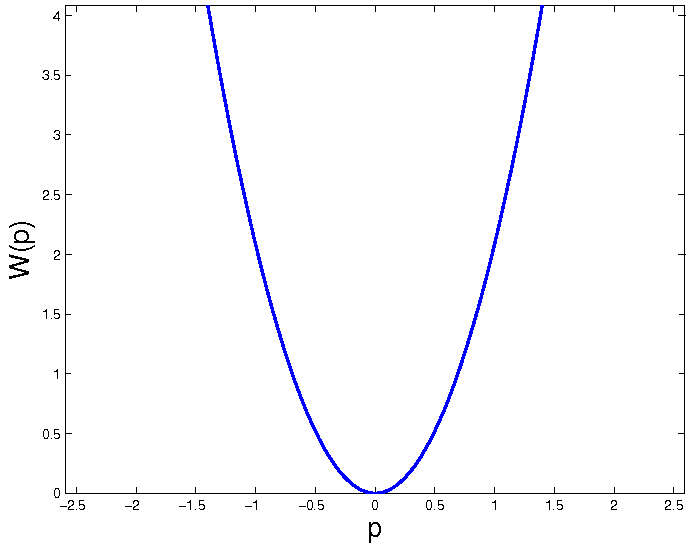}
\includegraphics[scale=0.3]{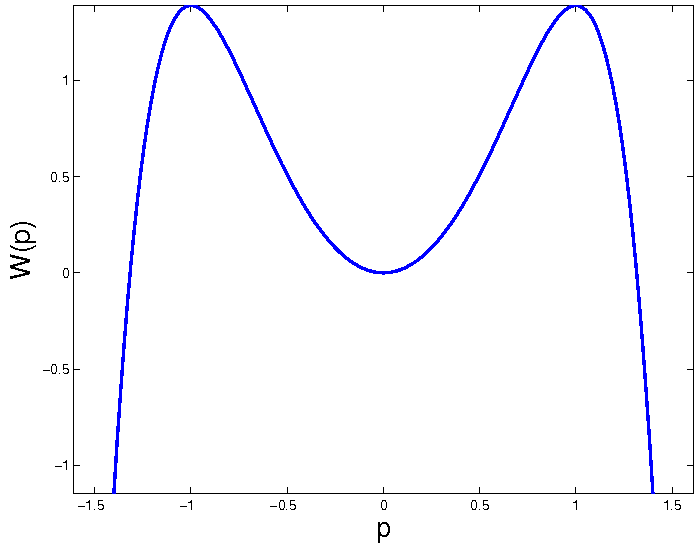}
\end{center}
\caption{{\protect\small Shape of the potentials $W(p)$ for $\alpha=0.01$ and $\beta=1$.
The first picture corresponds to the potential given by  $R^2$ gravity and the second one to the potential given by the  model suggested in \cite{s80}.
The dynamics  is very easy: one can imagine a "particle" moving under the action of the potential $W$
and losing energy.  For the first potential particles   come at early times from $|p|=\infty$ and ends at late time at  $p=0$ in an oscillating way. For the second potential there are
two unstable de Sitter solutions at $p=\pm(1/\beta)^{-1/4}$, so the "particle" could start at early times at these points and fall down
into the wedge of the potential  ending,  at late time, at  $p=0$ in
an oscillatory way. These are the only non-singular solutions. All the other orbits are singular at early and/or late times.
}}
\end{figure}

\vspace{1cm}

\begin{figure}[h]
\begin{center}
\includegraphics[scale=0.35]{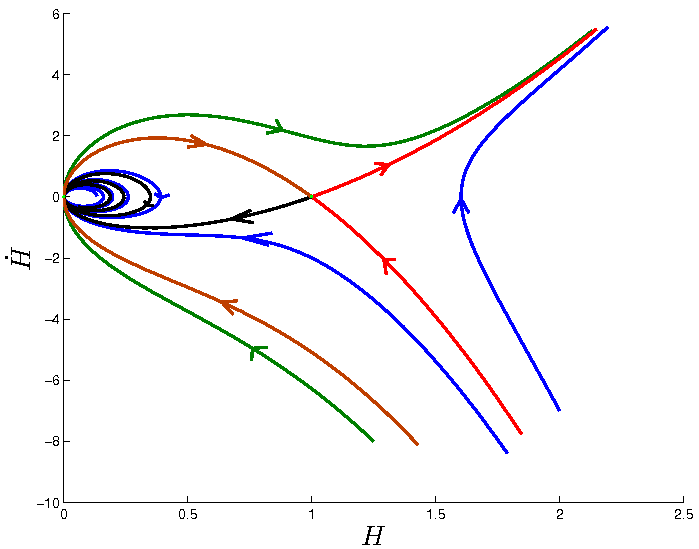}
\end{center}
\caption{{\protect\small Phase portrait for $\alpha=0.01$ and $\beta=1$ of the Starobinsky model. The unique non-singular solutions
are the de Sitter one which correspond to the saddle point $(1,0)$, which is the unique critical point of the system (painted brown), and the black curve that starts at the critical  point
and ends oscillating at $(0,0)$.
}}
\end{figure}

Note that what is really important in the Starobinsky model at late times,
 is the oscillatory behavior of the scale factor rather than
its amplitude, because at late times the period of oscillation of the scale factor is much shorter than the Hubble time, meaning that for a few oscillations the amplitude of the scale
factor can be considered constant.
This behavior can be thought of as oscillations of a decaying field called {\it scalaron} \cite{s80} that
creates light conformally coupled particles, which finally
thermalize yielding a hot Friedmann universe that matches with the Standard Model.

\section{Loop Quantum $R^2$ gravity}

We start this section showing that there exists a wide range of values of $\alpha$ and $\tilde{\rho}_c$ for which the $R^2$ LQC model does not have any singularity.
First at all, from the holonomy corrected Friedmann equation (\ref{fri}) we deduce that
\begin{eqnarray}
0\leq \tilde{\rho}\leq \tilde{\rho}_c \quad \mbox{and}\quad
-\sqrt{\frac{\tilde{\rho}_c}{12}}\leq \tilde{H}\leq \sqrt{\frac{\tilde{\rho}_c}{12}}.
\end{eqnarray}

On the other hand,
 equation (\ref{4}), applied to $R^2$ gravity,  leads to the positive potential
\begin{eqnarray}\label{pot}
 V(\tilde{\phi})=\frac{1}{8\alpha}\left(1-e^{-\sqrt{\frac{2}{3}}\tilde{\phi}} \right)^2,
\end{eqnarray}
 and thus, one also has
\begin{eqnarray}
0\leq (\tilde{\phi}')^2\leq 2\tilde{\rho}_c \quad \mbox{and}\quad
0\leq V(\tilde{\rho})\leq \tilde{\rho}_c.
\end{eqnarray}

Using the Raychaudhuri equation in LQC, $\tilde{H}'=-\frac{1}{2}(\tilde{\phi}')^2\left(1-\frac{2\tilde{\rho}}{\tilde{\rho}_c} \right)$, one deduces
that
\begin{eqnarray}
 |\tilde{H}'|\leq \frac{1}{2}(\tilde{\phi}')^2\leq \tilde{\rho}_c\Longrightarrow |\tilde{R}|\leq 7\tilde{\rho}_c.
\end{eqnarray}

Moreover,
the potential (\ref{pot})
 satisfies
\begin{eqnarray}
 \frac{\partial V(\tilde{\phi})}{\partial \tilde{\phi}}=\frac{1}{f_R(R)}\sqrt{\frac{V(\tilde{\phi})}{3\alpha}},
\end{eqnarray}
which means (see the second equation of (\ref{rel}))
\begin{eqnarray}
 R=f_R(R)\left(\tilde{R}+(\tilde{\phi}')^2\right)+\sqrt{\frac{2V(\tilde{\phi})}{\alpha}},
\end{eqnarray}
and thus,
\begin{eqnarray}
 R=\frac{1}{1-2\alpha(\tilde{R}+(\tilde{\phi}')^2)}\left(\tilde{R}+(\tilde{\phi}')^2+ \sqrt{\frac{2V(\tilde{\phi})}{\alpha}}\right).
\end{eqnarray}

From the bound $1-2\alpha(\tilde{R}+(\tilde{\phi}')^2)\geq 1-18\alpha\tilde{\rho}_c$ one easily deduces
\begin{eqnarray}
 |R|\leq \frac{1}{1-18\alpha\tilde{\rho}_c}\left(18\tilde{\rho}_c+ \sqrt{\frac{2\tilde{\rho}_c}{\alpha}}\right),
\end{eqnarray}
which is always bounded provided we choose $\alpha<\frac{1}{18\tilde{\rho}_c}$.

Finally, since $|R|$ is bounded, from the first equation of (\ref{rel}) one deduces that $|H|$ is bounded, and consequently
$|\dot{H}|=\frac{1}{6}\left|R-12H^2\right|$ is bounded, meaning that $R^2$ gravity  in LQC has no singularities.

In fact, as we will see, in any case there are singularities when one takes into account holonomy corrections. However, when
$8\alpha\tilde{\rho}_c>1$ the scalar curvature $R$ can achieve very large values. To show that, we have to perform a detailed
analysis in EF.

\subsection{$R^2$ LQC  in Einstein frame}
To perform a deeper analysis of the model we will work in EF, where the dynamical equations are simpler than in the JF one.
In fact,  when $f(R)=R+\alpha R^2$ equation
(\ref{fri}) becomes
\begin{eqnarray}\label{equacio}
 12\alpha\dot{R}H+6H^2(1+2\alpha R)-\alpha R^2=-\frac{\alpha^2[2\alpha(R^3+3\dot{R}^2)+R^2 ]^2}{2(1+2\alpha R)^4\tilde{\rho}_c}.
\end{eqnarray}

On the other hand,
in EF, the field $\tilde{\phi}$ satisfies the equation
\begin{eqnarray}\label{field}
 \tilde{\phi}''+3\tilde{H}\tilde{\phi}'+\frac{\partial V(\tilde{\phi})}{\partial \tilde{\phi}}=0,
\end{eqnarray}
where the potential $\tilde{\phi}$ is given by (\ref{pot}).

As we have already explained, due to the holonomy effects in EF the universe starts in the contracting phase  with zero energy, and the energy density increases as far as it catches up with the
critical value $\tilde{\rho}_c$, where the universe bounces and enters in the expanding phase.

Performing the change of variable $\sqrt{\frac{2}{3}}\tilde{\phi}=\ln\tilde{\psi}$, i.e. $\tilde{\psi}=f_R(R)=1+2\alpha R$ (essentially $\tilde{\psi}$ is like $R$), one gets
\begin{eqnarray}\label{field1}
 \tilde{\psi}''\tilde{\psi}-(\tilde{\psi}')^2+3\tilde{H}\tilde{\psi}'\tilde{\psi}+\frac{1}{6\alpha}\left(\tilde{\psi}-1\right)=0.
\end{eqnarray}

From equation (\ref{field1}) one can show that the orbits in the plane $(\tilde{\psi},\tilde{\psi}')$, are  symmetric with respect the axis $\tilde{\psi}'=0$
in the expanding and contracting phase, because equation
(\ref{field1})  remains invariant after performing the replacement $\tilde{t}\rightarrow-\tilde{t}$ and $\tilde{H}\rightarrow-\tilde{H}$. To be more precise, consider in the plane
$(\tilde{\psi},\tilde{\psi}')$,  a trajectory (a solution of (\ref{field1})) $\sigma_1(t)=(\tilde{\psi}(t),\tilde{\psi}'(t))$ in the contracting $\tilde{H}<0$ (resp. expanding $\tilde{H}>0$) phase. Then,
$\sigma_2(t)=(\tilde{\psi}(-t),-\tilde{\psi}'(-t))$ is a trajectory in the expanding $\tilde{H}>0$ (resp. contracting $\tilde{H}<0$) phase.

The energy density, using the new variables, is  given by
\begin{eqnarray}
 \tilde{\rho}=\frac{3}{4\tilde{\psi}^2}\left((\tilde{\psi}')^2 +\frac{1}{6\alpha}(\tilde{\psi}-1)^2 \right),
\end{eqnarray}
which means that $\tilde{H}$ vanishes at the point $(\tilde{\psi},\tilde{\psi}')=(1,0)$ and over the curve $\tilde{\rho}=\tilde{\rho}_c$, with equation
\begin{eqnarray}\label{curva}
\frac{(\tilde{\psi}')^2}{\frac{4\tilde{\rho}_c}{3(1-8\alpha\tilde{\rho}_c)}}+
\frac{(\tilde{\psi}-\frac{1}{1-8\alpha\tilde{\rho}_c})^2}{\frac{8\alpha\tilde{\rho}_c}{(1-8\alpha\tilde{\rho}_c)^2}}=1,
\end{eqnarray}
which produces an ellipse for $1-8\alpha\tilde{\rho}_c>0$, an hyperbola for $1-8\alpha\tilde{\rho}_c<0$ and a parabola for $1-8\alpha\tilde{\rho}_c=0$.
Note also that $(1,0)$ is the unique critical point
 corresponding to $\tilde{\rho}=0$, which means that all the orbits start and end at this point
 (the universe starts and ends at this point), and in the curve (\ref{curva}) the universe in EF bounces, because it corresponds to
$\tilde{\rho}=\tilde{\rho}_c$.

From the previous  analysis we can conclude that
the dynamics, working in EF, goes as follows: the universe
starts in the contracting phase $\tilde{H}<0$  oscillating around  the unique critical point $(1,0)$ and increasing the amplitude of oscillations, then it reaches
 the curve $\tilde{\rho}=\tilde{\rho}_c$ where it bounces and enters in the expanding phase
 $\tilde{H}>0$ coming back once again to $(1,0)$ in an oscillatory way (our analytical study is supported numerically in figure $4$).


\begin{figure}[h]
\begin{center}
\includegraphics[scale=0.4]{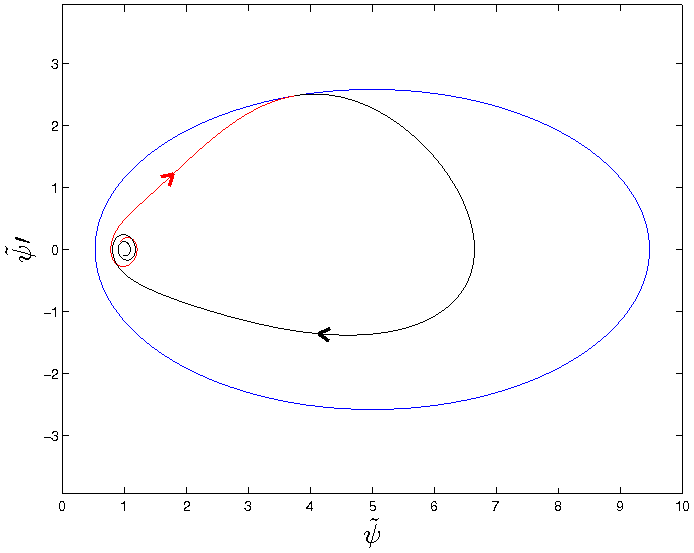}
\includegraphics[scale=0.4]{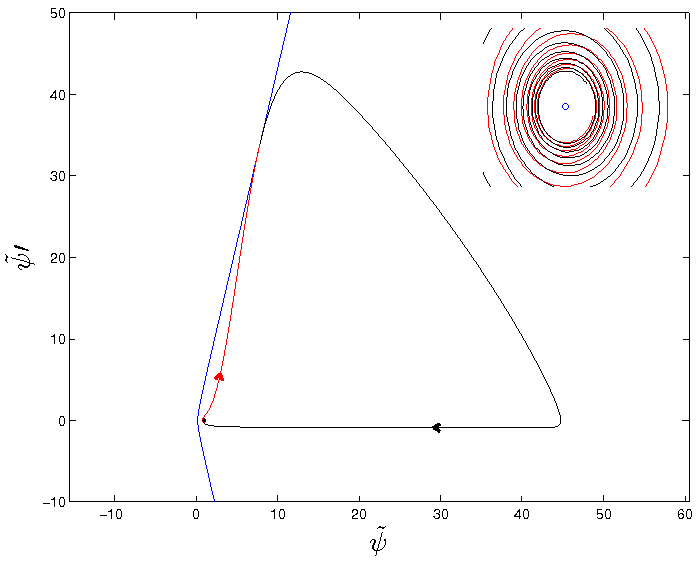}
\end{center}

\caption{{\protect\small In the first picture we have the phase space portrait of an orbit in EF for the case $1-8\alpha\tilde{\rho}_c>0$ ($\alpha=0.1$ and $\tilde{\rho}_c=1$).
The universe starts, in the contracting phase $\tilde{H}<0$, oscillating arround $(1,0)$ (red curve)
and  arriving to the ellipse defined by equation (\ref{curva}) (blue curve), where the universe bounces entering in the
expanding phase $\tilde{H}>0$ and coming back to  $(1,0)$ oscillating (black curve). In second picture we draw an orbit in EF for the case
$1-8\alpha\tilde{\rho}_c<0$ ($\alpha=0.1$ and $\tilde{\rho}_c=15$). The dynamics is similar, the only difference is that now the blue curve is an hyperbola. At the top of the picture
we have inserted and increased in size the oscillatory behavior around the critical point $(1,0)$.}}
\end{figure}

Two important remarks are in order:
\begin{enumerate}\item
 Strictly speaking, the phase portrait in the plane $(\tilde{\psi},\tilde{\psi}')$ shows the dynamics of two dynamical systems,
 because equation (\ref{field1}) defines two different differential equations, one with $\tilde{H}>0$ and the other one with $\tilde{H}<0$. Then, since we have two different
  autonomous  dynamical systems, at each point of the plane $(\tilde{\psi},\tilde{\psi}')$ two different orbits, one with   $\tilde{H}>0$ and the other one with $\tilde{H}<0$,
 cross.
\item
It is important to realize  that
the system does not contain singularities because all the orbits start and end at the critical point $(1,0)$.
In the case $1-8\alpha\tilde{\rho}_c>0$, the variables $\tilde{\psi}$ and  $\tilde{\psi}'$
move inside an ellipse (a compact domain) meaning that, in this case, all the quantities are bounded. Effectively,
inside the ellipse the quantities $\tilde{H}$, $\tilde{R}$, $\tilde{\phi}$, $\tilde{\phi}'$
and $\tilde{\psi}=1+2\alpha R$ are bounded. Consequently, it follows from (\ref{rel}) that $H$ is bounded. On the other hand, in the case $1-8\alpha\tilde{\rho}_c<0$
the variables $\tilde{\psi}$ and  $\tilde{\psi}'$ move inside an unbounded region delimited  by an hyperbola, meaning that there are
orbits where $\tilde{\psi}$, and
consequently the scalar curvature $R$,  achieve very large values, which never happens in the other case.

\end{enumerate}

 \subsubsection{Inflation in Einstein frame}

 The slow-roll parameters in EF are given by (see for example \cite{sl})
 \begin{eqnarray}\label{sr}
  \tilde{\epsilon}\equiv-\frac{\tilde{H}'}{\tilde{H}^2}\quad \mbox{and}\quad \tilde{\eta}\equiv \tilde{\epsilon}-\tilde{\delta}=2\tilde{\epsilon}-
  \frac{\tilde{\epsilon}'}{2\tilde{H}\tilde{\epsilon}},
 \end{eqnarray}
where $\tilde{\delta}=\frac{\tilde{\phi}''}{\tilde{H}\tilde{\phi}'}$.

Slow-roll dynamics requires $(\tilde{\phi}')^2\ll V(\tilde{\phi})$ and $\tilde{\phi}''\ll \tilde{H}\tilde{\phi}'$. Then, in the slow-roll phase the dynamical equations read
\begin{eqnarray}\label{slequation}
 \tilde{H}^2=\frac{V(\tilde{\phi})}{3}\left(1-\frac{V(\tilde{\phi})}{\tilde{\rho}_c}\right)\quad \mbox{and}\quad 3\tilde{H}\tilde{\phi}'+
 \frac{\partial V(\tilde{\phi}) }{\partial \tilde{\phi}}=0,
\end{eqnarray}
and thus, in this phase, the slow-roll parameters are approximately
\begin{eqnarray}\label{sr1}
  \tilde{\epsilon}\cong\frac{1}{2}
  \left( \frac{1}{V(\tilde{\phi})} \frac{\partial V(\tilde{\phi}) }{\partial \tilde{\phi}}\right)^2
  \frac{\left(1-\frac{2V(\tilde{\phi})}{\tilde{\rho}_c} \right)}{\left(1-\frac{V(\tilde{\phi})}{\tilde{\rho}_c} \right)^2}\quad
  \mbox{and}\quad \tilde{\eta}\cong
  \frac{1}{V(\tilde{\phi})} \frac{\partial^2 V(\tilde{\phi}) }{\partial \tilde{\phi}^2}\frac{1}{\left(1-\frac{V(\tilde{\phi})}{\tilde{\rho}_c} \right)}.
 \end{eqnarray}

For the potential given by $R^2$ gravity, i.e. for (\ref{pot}), slow-roll conditions ($|\tilde{\epsilon}|\ll 1$
and $|\tilde{\eta}\ll 1$) are only satisfied for large positive values of the field. In that case,  equation (\ref{sr1}) becomes
\begin{eqnarray}\label{sr2}
  \tilde{\epsilon}\cong\frac{4}{3}
  \frac{e^{-\sqrt{\frac{8}{3}}\tilde{\phi}}}{(1-e^{-\sqrt{\frac{2}{3}}\tilde{\phi}})^4}
  \frac{\left(1-\frac{(1-e^{-\sqrt{\frac{2}{3}}\tilde{\phi}})^2}{4\alpha\tilde{\rho}_c} \right)}{\left(1-\frac{(1-e^{-\sqrt{\frac{2}{3}}\tilde{\phi}})^2}{8\alpha\tilde{\rho}_c} \right)^2}
 \end{eqnarray}
  and
  \begin{eqnarray}
  \tilde{\eta}\cong
  \frac{4}{{3}}
  \frac{2e^{-\sqrt{\frac{8}{3}}\tilde{\phi}}-e^{-\sqrt{\frac{2}{3}}\tilde{\phi}}}{(1-e^{-\sqrt{\frac{2}{3}}\tilde{\phi}})^2}
  \frac{1}{\left(1-\frac{(1-e^{-\sqrt{\frac{2}{3}}\tilde{\phi}})^2}{8\alpha\tilde{\rho}_c} \right)}.
 \end{eqnarray}

 To calculate
 inflation ends, the values of the slow-roll parameters must be of the order $1$, which happens, for positive values of the field  $\tilde{\phi}$, when it satisfies the equation
 \begin{eqnarray}
  \frac{e^{-\sqrt{\frac{2}{3}}\tilde{\phi}}}{(1-e^{-\sqrt{\frac{2}{3}}\tilde{\phi}})^2}\cong\frac{\sqrt{3}}{2},
 \end{eqnarray}
whose solution is
 \begin{eqnarray}
  \tilde{\phi}_{end}=-\sqrt{\frac{3}{2}}\ln\left(\frac{1+\sqrt{3}-\sqrt{2\sqrt{3}+1}}{\sqrt{3}} \right)>0.
 \end{eqnarray}

And to calculate
 the number of e-folds that the scale factor increases during the period of inflation
 \begin{eqnarray}\tilde{N}\equiv \int_{\tilde{t}_{in}}^{\tilde{t}_{end}}\tilde{H}d\tilde{t}=
 \int_{\tilde{\phi}_{in}}^{\tilde{\phi}_{end}}\frac{\tilde{H}}{\tilde{\phi}'}d\tilde{\phi},
 \end{eqnarray}
we have to use the slow roll equations (\ref{slequation}) obtaining
 \begin{eqnarray}\tilde{N}\cong
 \int_{\tilde{\phi}_{end}}^{\tilde{\phi}_{in}}
 \frac{V(\tilde{\phi})}{\frac{\partial V(\tilde{\phi}) }{\partial \tilde{\phi}}}
 \left(1-\frac{V(\tilde{\phi})}{\tilde{\rho}_c}\right)
 d\tilde{\phi}
 \end{eqnarray}

 In the case of our potential (\ref{pot}), the final number of e-folds is approximately
 \begin{eqnarray}
  \tilde{N}\cong \frac{3}{4}e^{\sqrt{\frac{2}{3}}\tilde{\phi}_{in}}.
 \end{eqnarray}


 On the other hand, for a given value of $\tilde{N}$ the slow-roll parameters  are:
 \begin{eqnarray}
  \tilde{\epsilon}\cong \frac{3}{4\tilde{N}^2}\frac{\left(1-\frac{1}{4\alpha\tilde{\rho}_c} \right)}{\left(1-\frac{1}{8\alpha\tilde{\rho}_c} \right)^2}\quad \mbox{and}\quad
  \tilde{\eta}\cong -\frac{1}{\tilde{N}}\frac{1}{\left(1-\frac{1}{8\alpha\tilde{\rho}_c} \right)}.
 \end{eqnarray}

 With these values,  the spectral index of scalar perturbations, namely $\tilde{n}_s$, and the ratio of tensor to scalar perturbations, namely $\tilde{r}$, are approximately
 \begin{eqnarray}
  \tilde{n}_s\cong 1-6\tilde{\eta}+2\tilde{\eta}\cong 1 -\frac{2}{\tilde{N}}\frac{1}{\left(1-\frac{1}{8\alpha\tilde{\rho}_c} \right)}, \quad
  \tilde{r}\cong 16\tilde{\epsilon}\cong \frac{12}{\tilde{N}^2}\frac{\left(1-\frac{1}{4\alpha\tilde{\rho}_c} \right)}{\left(1-\frac{1}{8\alpha\tilde{\rho}_c} \right)^2},
 \end{eqnarray}
which coincide, when  holonomy corrections are disregarded, i.e. when $\tilde{\rho}_c\rightarrow \infty$, with the values obtained in \cite{bno}.

A very important remark is in order: The latest Planck data gives for the spectral index the approximate value $\tilde{n}_s=0.9603\pm 0.0073$. If one disregards the loop corrections, to achieve
the value $0.96$ one has to take  $\tilde{N}=50$ e-folds, which does not give enough inflation to solve the flatness and horizon problems. However, if one takes into account
holonomy corrections, for the values $8\alpha \tilde{\rho}_c\cong 6$ and $\tilde{N}= 60$ (the minimum number of e-folds required to solve the horizon and flatness problems)
one obtains  the desired result. Moreover, for these same values one obtains $\tilde{r}= 0.0031$, which satisfies the current bound $\tilde{r}< 0.11$.

To be more precise, if one disregards loop corrections, $60$ e-folds are only achieved when $0.9666\leq\tilde{n}_s\leq \tilde{n}_{s,max}=0.9676$,
in fact for $\tilde{n}_s=0.9676$ one obtains $61.72$ e-folds, which means that, in this model without corrections, it is impossible for the universe to inflate more that $61.72$ e-folds. However,
including
loop quantum effects one easily achives a greater number of e-folds; for example, for $\tilde{n}_s=0.9676$ one obtains $70$ e-folds choosing $8\alpha\tilde{\rho}_c\cong 8.46$.
To sum up, we have shown that
loop quantum corrections
could be essential to match correctly $R^2$ inflation with the current observational data.

\subsection{$R^2$ LQC in Jordan frame}
To study the dynamics in the JF from the results obtained in the EF,
 we look for the points in the space $(\tilde{\psi},\tilde{\psi}')$ where the universe could  bounce in the JF, i.e., we look for the points  where $H=0$. Since,
${H}=\sqrt{f_R(R)}\left(\tilde{H}-\frac{1}{\sqrt{6}}\tilde{\phi}' \right)$, one has to solve the equation
$\tilde{H}^2=\frac{1}{{4}}\frac{(\tilde{\psi}')^2}{\tilde{\psi}^2}$, which gives, for  $\tilde{\psi}>1$ the following curve
\begin{eqnarray}\label{curva1}
\frac{(\tilde{\psi}')^2}{\frac{\tilde{\rho}_c}{12(1-\sqrt{8\alpha\tilde{\rho}_c})}}+
\frac{\left(\tilde{\psi}-\frac{1-\sqrt{2\alpha\tilde{\rho}_c}}{1-\sqrt{8\alpha\tilde{\rho}_c}}\right)^2}{\frac{2\alpha\tilde{\rho}_c}
{\left(1-\sqrt{8\alpha\tilde{\rho}_c}\right)^2}}=1,
\end{eqnarray}
which, as in EF, produces an ellipse for $1-8\alpha\tilde{\rho}_c>0$, an hyperbola for $1-8\alpha\tilde{\rho}_c<0$ and a parabola for $1-8\alpha\tilde{\rho}_c=0$. And,
for $0<\tilde{\psi}<1$ the curve is
\begin{eqnarray}\label{curva2}
\frac{(\tilde{\psi}')^2}{\frac{\tilde{\rho}_c}{12(1+\sqrt{8\alpha\tilde{\rho}_c})}}+
\frac{\left(\tilde{\psi}-\frac{1+\sqrt{2\alpha\tilde{\rho}_c}}{1+\sqrt{8\alpha\tilde{\rho}_c}}\right)^2}{\frac{2\alpha\tilde{\rho}_c}
{\left(1+\sqrt{8\alpha\tilde{\rho}_c}\right)^2}}=1,
\end{eqnarray}
which is always an ellipse. Then, when in EF the orbits in the plane $(\tilde{\psi},\tilde{\psi}')$ reach those curves, the universe in the JF could bounce. To assure that it bounces the equation $\tilde{H}=\frac{\tilde{\psi}'}{2\tilde{\psi}}$ must be satisfied.

Now we are ready to explain the dynamics in JF from the results already obtained in EF: In EF the dynamics starts in the contracting phase
and ends in the expanding one at the critical point $(\tilde{\psi},\tilde{\psi}')=(1,0)$. From the relation  between both frames
\begin{eqnarray}\label{rel1}
H=\sqrt{\tilde{\psi}}\left(\tilde{H}-\frac{\tilde{\psi}'}{2\tilde{\psi}}\right),\quad \dot{H}=\frac{\tilde{\psi}'}{2}\left(\tilde{H}-\frac{\tilde{\psi}'}{2\tilde{\psi}}\right)+
\tilde{\psi}\left(\tilde{H}'-
\frac{1}{2}\left(\frac{\tilde{\psi}'}{\tilde{\psi}}\right)'\right),
\end{eqnarray}
which is obtained from the first equation of (\ref{rel}) and its derivative, one deduces that, in JF, the universe starts and ends at $(H=0,\dot{H}=0)$. Note that to calculate
explicitly $\dot{H}$ one has to use the Raychaudhuri equation $\tilde{H}'=-\frac{3}{4}\left(1-\frac{2\tilde{\rho}}{\tilde{\rho}_c}\right)\left(\frac{\tilde{\psi}'}{\tilde{\psi}}\right)^2$
and the field equation (\ref{field1}).
Moreover, since in EF the orbits of the system at early and late times oscillate  around the point $(\tilde{\psi},\tilde{\psi}')=(1,0)$, crossing many times
the curves (\ref{curva1}) and (\ref{curva2}) one can conclude that
in JF the orbits of the system at early times oscillate around the point $(H,\dot{H})=(0,0)$ meaning that the universe makes small bounces many times, and
when it leaves this  oscillatory regime  it enters in the contracting
phase and bounces (in EF when the orbit reach the curve (\ref{curva})) to enter in the expanding phase, where the universe inflates and finally, at late times, it goes asymptotically
to the critical point $(0,0)$ in an oscillating way, that is, bouncing again many times.

Note that this behavior is completely different from the one obtained disregarding holonomy corrections where, in JF, as we have already seen in section IV, the universe never bounces
and is singular at early times. Moreover,
it is important to remark that the holonomy corrected  equation (\ref{equacio}) is not singular at $H=0$, and thus the orbits can cross the axis $H=0$, which allows the
universe to bounce. Of course, that does not happen in  classical $R^2$ gravity  where the corresponding dynamical equation (eq. (\ref{sta})) is not defined at $H=0$.


Numerically, the dynamics in the plane $(H,\dot{H})$ is  easily derived
via (\ref{rel1})
from the one in EF, which is very simple as we have already shown. In figure 5, we have depicted in the plane $(H,\dot{H})$ the orbits depicted  in figure 4.


\begin{figure}[h]
\begin{center}
\includegraphics[scale=0.45]{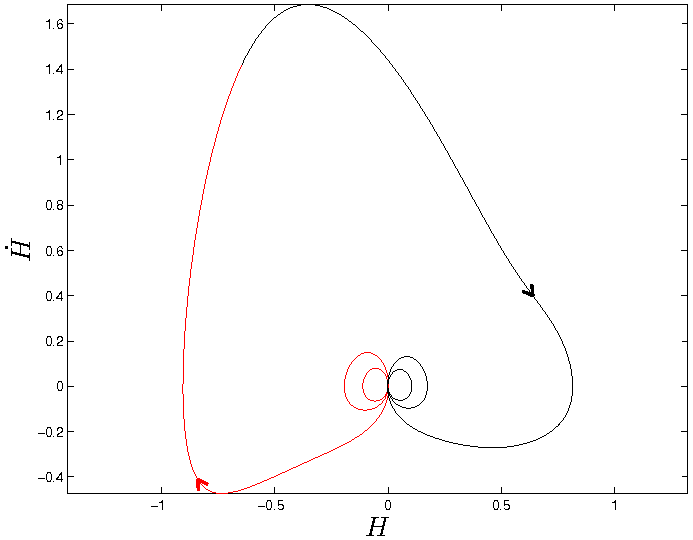}
\includegraphics[scale=0.45]{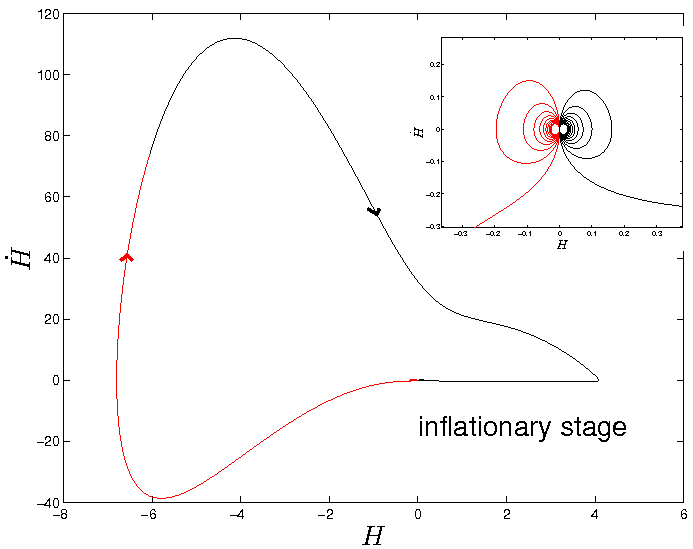}
\end{center}

\caption{{\protect\small In the first picture we have the phase space portrait of an orbit in JF for the case $1-8\alpha\tilde{\rho}_c>0$ ($\alpha=0.1$ and $\tilde{\rho}_c=1$).
The universe starts oscillating around $(0,0)$
then enters in  the contracting phase $(H<0)$ and  bounces entering in the
expanding phase ${H}>0$  coming back to  $(0,0)$ oscillating. In the second picture we draw an orbit in JF for the case
$1-8\alpha\tilde{\rho}_c<0$ ($\alpha=0.1$ and $\tilde{\rho}_c=15$). The dynamics is similar to that described in the other picture, but there is enough inflation here in the
expanding phase.}}
\end{figure}

Note also that  equation $(\ref{field1})$ defines two different dynamical systems, which means that in  the plane $(H,\dot{H})$, two different
orbits, one with  $\tilde{H}>0$ and the other one with $\tilde{H}<0$,  cross at each point. Moreover, the invariance of the equation
(\ref{field1}) with respect to the the replacement $\tilde{t}\rightarrow-\tilde{t}$ and $\tilde{H}\rightarrow-\tilde{H}$, means that the phase portrait in the plane
$(H,\dot{H})$ has a symmetry with respect the axis $H=0$. More precisely, given a piece of an orbit with $\tilde{H}>0$ (resp. $\tilde{H}<0$) in EF, there is  a symmetric
piece with respect to the axis $H=0$,
 of an orbit with $\tilde{H}<0$ (resp. $\tilde{H}>0$) in EF.

A final remark is in order: In JF, the dynamics of our extension of LQC to $R^2$ gravity in the vaccum (we do not have considered any scalar field),  is given by equation
(\ref{equacio}) which is a second order differential equation on $H$, mathematically meaning that is a first order differential system in variables $(H,\dot{H})$.
Working in EF the dynamics is depicted by
equation (\ref{field}) which is also a second order  in $\tilde{\phi})$ (note that from the holonomy corrected Friedmann equation $\tilde{H}$ is merely a function of
$\tilde{\phi}$ and $\tilde{\phi}'$, in the same way as in standard LQC), meaning that the dynamics is given by a first order differential system in variables
($\tilde{\phi},\tilde{\phi}'$).

The same happens in $f(R)$ gravity ``\`a la Palatini'' where the connection is a free variable (see for instance \cite{olmo}) and in teleparallel $f(T)$ gravity \cite{aho},
when the stress-tensor is depicted by an
scalar field $\phi$, because in both cases the corresponding modifed Friedmann equation relates the Hubble parameter with the energy density, i.e., the Hubble parameter ia a
function of ${\phi}$ and $\dot{\phi}$, meaning that the dinamics is given by the conservation equation
\begin{eqnarray}
 \ddot{\phi}+3H\dot{\phi}+\frac{\partial V}{\partial \phi}=0,
\end{eqnarray}
which is a second order differential equation in $\phi$.

However, if one considers  standard $f(R)$ gravity or $f(R)$ LQC, i.e., if the connection is fixed to be the Levi-Civit\`a one,  coupled with a scalar field $\phi$, the number of degrees
of freedom will increasse because appart from the modified
Friemann equation in $f(R)$ gravity, which is second order in $H$, one has to consider the conservation equation, which is second order in $\phi$, meaning that one will
have a first order differential system in the plane $(H,\dot{H},\phi,\dot{\phi})$.

 \section{Inflation and Dark Energy in $R^2$ LQC}
 Some time ago the unification of the early time inflation with late time Dark Energy (DE) in frames of modified gravity was proposed (\cite{no03}). Later,
 several improved models containing Dark Energy (DE) have been suggested to unify inflation with the current acceleration of the universe.  In this work,  the idea is to add to $R^2$ gravity
 a correction $g(R)$ given a model of the form $f(R)=R+\alpha R^2+g(R)$ that takes into account the accelerated expansion of the universe and passes the Solar system tests. Two of the
 best regarded examples
 of these
 corrections are:
 $g(R)=\lambda(e^{-bR}-1)$ being $\lambda$ and $b$ positive constants \cite{cenosz}, and
 $g(R)=-m^2\frac{c_1\left(R/m^2 \right)^n}{c_2\left(R/m^2 \right)^n+1}$, where $n>0$ and $c_1, c_2$ are dimensionless parameters \cite{hs}.

 The problem with this kind of  models is that they lead to  very complicated potentials in EF,  complicating considerably their extension to LQC.
 Moreover, it is nearly impossible to perform a
 detailed analytical study and it is not evident how to perform numerical computations. For this reason in order to deal with DE we will consider the simplest model:
  we will add an small cosmological constant to our model, i.e., we will consider the $f(R)=R+\alpha R^2-2\Lambda$ model.

  When one does not take into
 account holonomy corrections, the system after the change $p=H^2$ has the same form as (\ref{sis}) but with the potential $W(p)=\frac{p^2}{48\alpha}+\frac{\Lambda}{144\alpha p^2}$.
 This potential satisfies $V(0)=V(\infty)=\infty$ meaning that the dynamics can be restricted to positive values of $p$. The potential only has a minimum at the point
 $p=\left(\frac{\Lambda}{3}\right)^{1/4}$ (de Sitter solution), and thus at late times all the solutions go asymptotically to this point oscillating around it. Moreover,
 the inflationary solution given in (\ref{infla}) is also an attractor when  the cosmological constant is taken into account. Finally, it is easy to show that the solutions are
 singular at early times. When a cosmological constant is considered, there are two kind of solutions:
 the ones that, as in $R^2$ gravity without cosmological constant, are given by $p(t)=\sqrt{\frac{1}{2(t-\bar{t})}}$, and the other ones given by
$p(t)=\left(\frac{\Lambda}{36\alpha}(t-\bar{t})\right)^{1/6}$, which vanish at $t=\bar{t}$ but have divergent scalar curvature.

Incorporating the cosmological constant to the EF model we have obtained the following potential $V(\tilde{\phi})=\frac{1}{8\alpha}\left(1-e^{-\sqrt{\frac{2}{3}}\tilde{\phi}} \right)^2+
 \Lambda e^{-\sqrt{\frac{8}{3}}\tilde{\phi}}$, which has a minimum at $\tilde{\phi}_{min}=\sqrt{\frac{3}{2}}\ln (1+8\alpha\Lambda)$. That means that, at late times in
 the plane $(\tilde{\phi},\tilde{\phi}')$ of EF, all the solutions oscillate around $\tilde{Q}_{min}\equiv (\tilde{\phi}_{min},0)$. When we introduce Loop Quantum effects
 in EF, the orbits will oscillate initially around $\tilde{Q}_{min}$ in the contracting phase, i.e.,  $\tilde{H}<0$. In fact, $\tilde{Q}_{min}$  in the contracting
 phase corresponds to the anti de Sitter solution $\tilde{H}_-=-\sqrt{\frac{V(\tilde{\phi}_{min})}{3}\left(1-\frac{V(\tilde{\phi}_{min})}{\tilde{\rho}_c} \right)}$,
 where $V(\tilde{\phi}_{min})=\frac{\Lambda}{8\alpha\Lambda+1}$ is the minimum value of the potential. After leaving the anti de Sitter phase  the orbits move into the contracting
 phase before bouncing and entering in the expanding one where the universe inflates, and finally oscillate asymptotically to the de Sitter solution
 $\tilde{H}_+\equiv -\tilde{H}_-$.

In JF, the dynamics is very similar: the universe starts oscillating around  the anti de Sitter solution $H_-=\sqrt{8\alpha\Lambda+1}\tilde{H}_-$, after leaving the
anti de Sitter phase moves in the contracting phase $H<0$, which it leaves bouncing, enters the expanding phase where it inflates and finally, at late times, it oscillates around
 the de Sitter
solution $H_+=\sqrt{8\alpha\Lambda+1}\tilde{H}_+$. This oscillatory behavior at late  times is essential, because it excites the light fields coupled with gravity
that will re-heat the universe \cite{mms,w,s80}, yielding a hot universe that matches with the $\Lambda$CDM model.

\vspace{1cm}

\section{Conclusions}
We have introduced holonomy corrections to $R^2$ gravity in order to avoid early time singularities that appear in this model.
We have performed a detailed analytical and numerical analysis which shows that the new model is not singular due to
 the quantum geometric corrections (holonomy corrections) coming from the discrete nature of space-time assumed in LQC. The new model
 is more involved than the original one. For this reason, in order to understand the dynamics in  JF, a previous analysis must
 be performed in EF, where the dynamical equations greatly simplify. This allows us to perform a detailed study of its dynamics, what is essential in order
 to have a global idea  of the system in JF. From this analysis we conclude that, when quantum geometric corrections are taken into account, the universe evolves from the
 contracting phase to the expanding one through a big bounce, and when it enters in the expanding phase, as in the classical model, it inflates in such a way that,
 these holonomy corrections lead to theoretical predictions that match  correctly with current observational data.
Finally, to remark that
it would be interesting to study different versions of F(R) gravity, for instance, with several power-law type terms in order to understand how such theories which normally
do not support the inflation behave in LQC approach.

\vspace{1cm}
{\bf Acknowledgments:}
This investigation has been
supported in part by MINECO (Spain), project MTM2011-27739-C04-01, MTM2012-38122-C03-01,
 by AGAUR (Generalitat de Ca\-ta\-lu\-nya),
contracts 2009SGR 345 and 994, and by
project TSPU-139 of Russ. Min. of Education and Science (SDO).

\end{document}